\documentclass[aip,reprint]{revtex4-1}
\pdfoutput=1
\usepackage{graphicx,psfrag,epsf,epsfig}
\usepackage{dcolumn}
\usepackage{bm,bbm}
\usepackage{pstricks,pst-plot}     
\usepackage{latexsym}
\usepackage{mathrsfs}
\usepackage{amsmath}
\usepackage{amssymb}
\usepackage{textcomp}

\usepackage{balance}

\begin{document}


\title{Impact of built-in fields and contact configuration on the characteristics of ultra-thin GaAs solar cells} 



\author{Urs Aeberhard}
\email[]{u.aeberhard@fz-juelich.de}
\affiliation{IEK-5 Photovoltaik, Forschungszentrum J\"ulich, D-52425 J\"ulich, Germany}


\date{\today}

\begin{abstract}
We discuss the effects of built-in fields and contact configuration on the photovoltaic characteristics of ultrathin GaAs solar cells. The investigation is based on advanced quantum-kinetic simulations reaching beyond the standard semi-classical bulk picture concerning the consideration of charge carrier states and dynamics in complex potential profiles. The thickness dependence of dark and photocurrent in the ultra-scaled regime is related to the corresponding variation of both, the built-in electric fields and associated modification of the density of states, and the optical intensity in the films. Losses in open-circuit voltage and short-circuit current due to leakage of electronically and optically injected carriers at minority carrier contacts are investigated for different contact configurations including electron and hole blocking barrier layers. The microscopic picture of leakage currents is connected to the effect of finite surface recombination velocities in the semi-classical description, and the impact of these non-classical contact regions on carrier generation and extraction is analyzed. 
\end{abstract}


\maketitle 

Solar cells based on high-quality crystalline materials with strong optical absorption, such as GaAs, are known to provide very high single junction efficiencies \cite{kayes:11,wang:13_jpv,steiner:13}. However, the fabrication of ordinary solar cells based on such materials is expensive and limits the commercial viability of the technology. Recently, ultra-thin high-efficiency absorber architectures were proposed as one way to mitigate the problem of high material and fabrication costs \cite{wang:13,massiot:14,yang:14,vandamme:15}. To compensate for the reduced absorber thickness, the absorption in these solar cells is enhanced by nanophotonic light-trapping structures. In this way, almost complete absorption of the solar spectrum could be achieved over a wide range of photon frequencies  with absorber thicknesses below 100 nm. 

For the conventional single-junction $p$-$i$-$n$ architectures exhibiting a doping-induced built-in potential of the order of the band gap energy at short-circuit conditions, such ultra-scaled absorber dimensions amount to very strong built-in fields even at the operating point of the solar cell. Hence, the validity of the assumption of bias-independent flat-band bulk material properties conventionally used in semiclassical solar cell simulation models becomes increasingly questionable for these devices. A further aspect of ultra-thin absorber architectures where the conventional treatment is likely to miss the physical picture concerns the electron and hole blocking layers introduced to increase the carrier selectivity by reducing leakage of minority carriers. The doped barrier layers introduce interfacial regions where charge carriers cannot be described properly within a classical picture, and which become increasingly relevant as the bulk absorber volume decreases.      

\begin{figure}[t!]
\begin{center}
\includegraphics[width=0.4\textwidth]{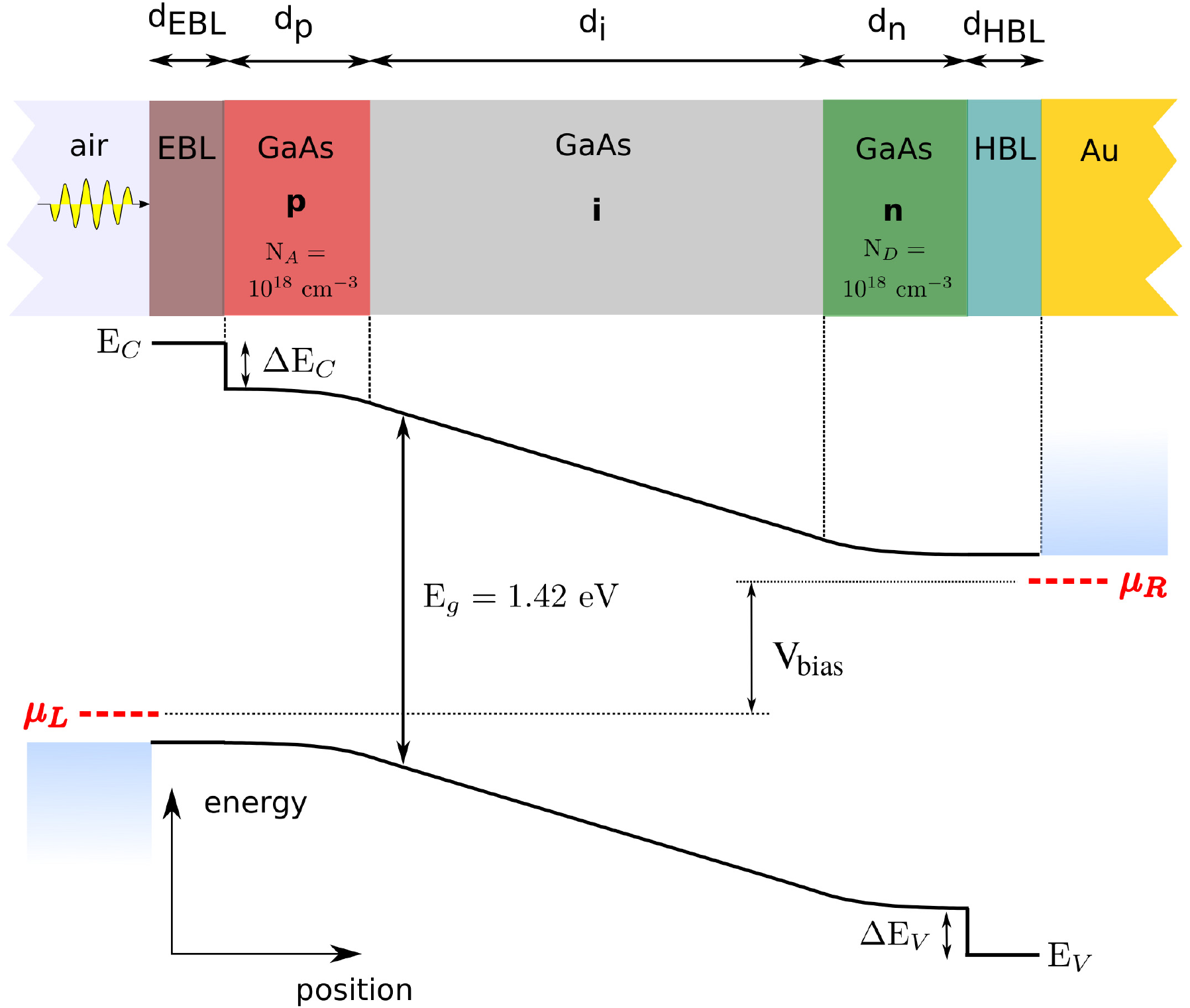}
\caption{Model system used in the simulations, corresponding to an ultrathin $p$-$i$-$n$ photodiode. The chemical potentials $\mu_{\textrm{L/R}}$ at left and right contacts are split by the applied bias voltage $V_{\textrm{bias}}$. The electron (hole) blocking layer EBL (HBL) prevents carrier leakage via band offset $\Delta E_{\textrm{C}}$ ($\Delta E_{\textrm{V}}$). The gold reflector is only used for the optical simulation.  \label{fig:structure_bands}}
\end{center}
\end{figure}

In order to assess the validity of the semiclassical bulk picture in ultra-scaled photovoltaic architectures, one needs to resort to a theoretical framework where the simplifying assumptions of the conventional approach are relaxed. A theoretical framework that is suitable for this purpose is the quantum-kinetic non-equilibrium Green's function formalism (NEGF) \cite{ae:jcel_11}. This approach is able to capture any deviation from bulk-like behavior induced in the electronic properties by strong fields or heterostructure potentials \cite{ae:jpe_14} and in the optical properties by nanophotonic structures \cite{ae:oqel_14}. Recently, the NEGF picture was implemented for ultra-thin GaAs solar cell devices at the radiative limit and revealed substantial deviations from the bulk behavior provided by the semiclassical description \cite{ae:jpv_16}.  

In this letter, we investigate the thickness dependence of the photovoltaic characteristics due to the variation of absorption and emission with built-in fields and optical intensity. We further consider the effects of non-classical regions induced by the barrier potentials of electron an hole blocking layers mediating enhanced carrier selectivity of contacts.  

\begin{figure}[t]
\begin{center}
\includegraphics[width=0.45\textwidth]{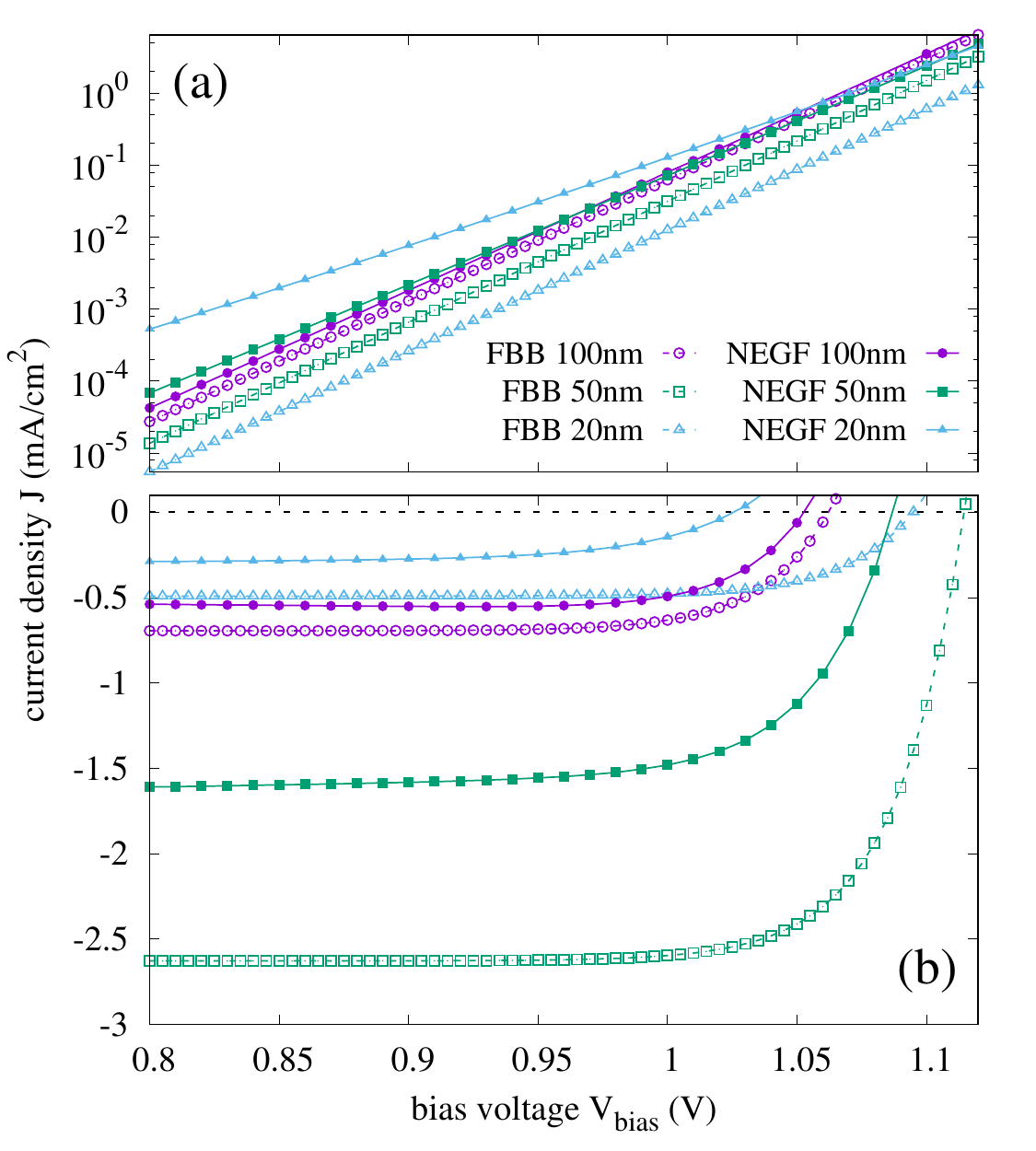}
\caption{(a) Dark current-voltage characteristics for absorber thicknesses of 100~nm, 50~nm and 20~nm, computed using the quantum-kinetic approach (NEGF, filled symbols) and the semiclassical flat-band bulk model (FBB, open symbols). The increase of the built-in potential with reduced absorber thickness leads to a strong enhancement of dark current due to recombination between field-induced tail states. (b) Characteristics under monochromatic illumination ($E_{\gamma}=1.44$ eV, $I_{\gamma}=0.1$ kW/m$^2$) for the same devices and simulation approaches as in (a). The effect of increasing fields on the absorption is partially masked by the strong fluctuation of the optical intensity with thickness variation. \label{fig:td} }
\end{center}
\end{figure}

For the semi-classical simulation of generation, transport and recombination of charge carriers in ultra-thin absorber devices, the conventional 1D drift-diffusion-Poisson solver ASA (Advanced Semiconductor Analysis, TU Delft) is used. The generation rate is  computed locally from the differential absorptance as a function of the intensity of the transverse electromagnetic field as given by the built-in transfer matrix method (TMM) and of the local absorption coefficient. The coherent wave approach of the TMM enables the consideration of optical resonator modes at sub-wavelength absorber thicknesses. 
Recombination is restricted to the fundamental radiative process as described  by the detailed-balance approach in terms of the local absorption coefficient and the black body radiation flux \cite{roosbroeck:54}. 

The quantum-kinetic simulations are performed with an in-house developed NEGF-Poisson solver that considers the non-equilibrium quantum statistical mechanics of open electronic systems featuring complex nanoscale potentials and in the presence of electron-photon and electron-phonon interactions \cite{ae:prb_08}. For the electronic structure of the bulk absorber materials, two decoupled single orbital tight-binding bands are used, with parameters obtained via the analogy to the effective mass approximation (EMA) \cite{lake:97}. The photogeneration is described in terms of coupling to the classical transverse field as obtained again from the TMM, while for the spontaneous emission, coupling to the photon density of states of the absorber as encoded in the slab photon Green's function is considered \cite{ae:prb89_14}.     

\begin{figure}[t]
\begin{center}
\includegraphics[width=0.45\textwidth]{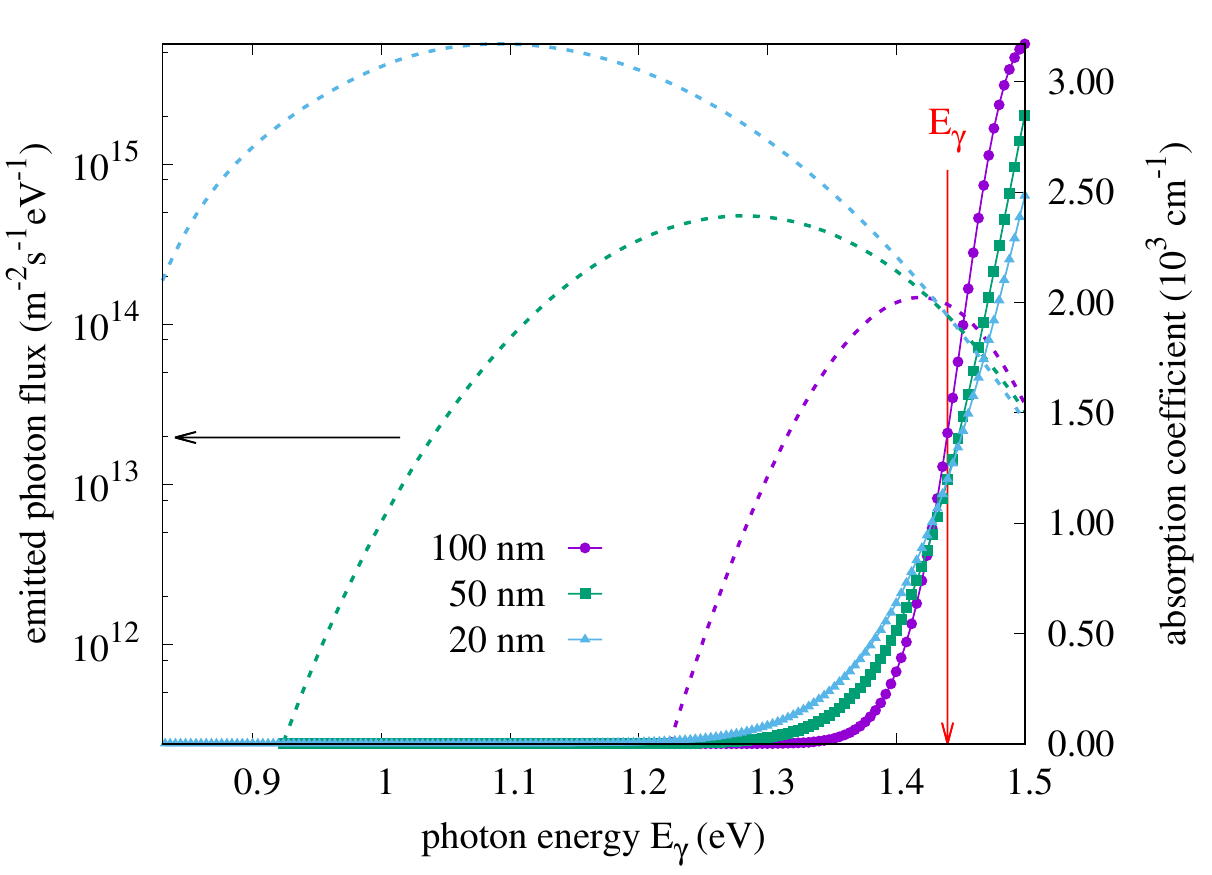}
\caption{Absorption coefficient and spectral emission rate of the $p$-$i$-$n$ diode at $V_{\textrm{bias}}=0.8$ V, for absorber thicknesses of 100~nm, 50~nm and 20~nm, respectively. While the value of the absorption coefficient is almost constant for the photon energy $E_{\gamma}=1.44$ eV of the monochromatic illumination, the emission rate exhibits a pronounced red shift and broadening with shrinking absorber thickness owing to the field-induced tailing of the joint density of states. \label{fig:absem_td} }
\end{center}
\end{figure}
 
In order to maximize the comparability of the macroscopic semiclassical approach with the microscopic quantum formalism, in addition to relying on the same optical model (TMM), the material parameters in the semi-classical approach were extracted from the two band EMA model of the bulk electronic structure used in the microscopic picture. The numerical values of the parameters used can be found in Ref.~\onlinecite{ae:jpv_16}. Since this simple band structure deviates considerably from the experimental data at energies far from the band edges, the investigation is focused on the device behavior at excitation energies close to the band gap. 


\begin{figure}[t]
\begin{center}
\includegraphics[width=0.4\textwidth]{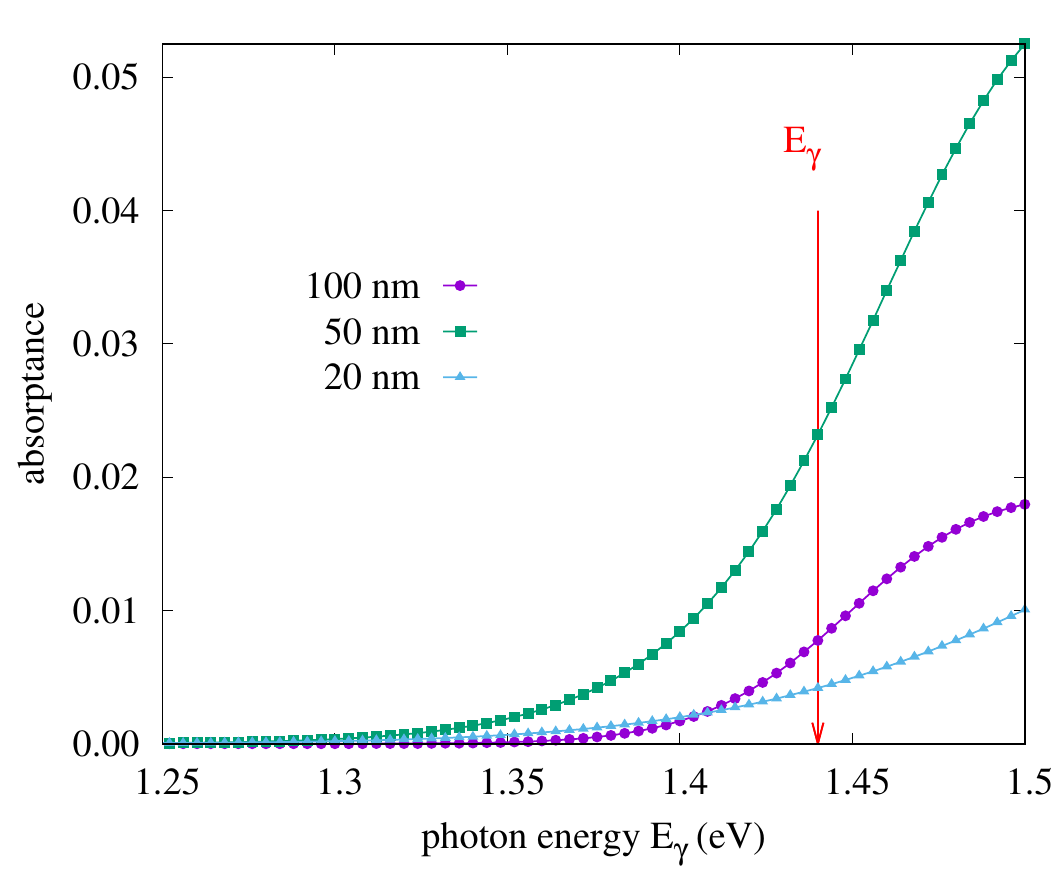}
\caption{Absorptance of the $p$-$i$-$n$ structure at $V_{\textrm{bias}}=0.8$ V for the different absorber thicknesses, exhibiting a strongly non-monotonous behavior reflected in the short-circuit current of the characteristics in Fig.~\ref{fig:td}(b), which originates in the strong variation of the optical intensity for the different configurations. \label{fig:absorpt_td} }
\end{center}
\end{figure}

In a recent publication, the qualitative and quantitative discrepancies between the drift-diffusion picture assuming flat band bulk material and the NEGF simulations of photovoltaic device characteristics were investigated for a simple $p$-$i$-$n$ structure as shown in Fig.~\ref{fig:structure_bands}, with $N_{A,D}=10^{18}$ cm$^{-3}$, $d_{\textrm{n,p}}$=20~nm, $d_{\textrm{i}}$=60 ~nm and perfectly selective contacts ($\Delta E_{\textrm{C,V}}\rightarrow \infty$) \cite{ae:jpv_16}.  It was found that differences originate mostly in the effects of the strong built-in fields on absorption and emission (Franz-Keldysh effect). With further thickness reduction, the discrepancies between semiclassical flat-band bulk (FBB) and quantum-kinetic (NEGF) descriptions that are related to field effects -- absent in the FBB -- increase due to growing built-in potentials, as displayed in Fig.~\ref{fig:td}. In subfigures (a) and (b), the current-voltage characteristics are shown for the dark and under illumination with monochromatic light of photon energy $E_{\gamma}=1.44$ eV and intensity $I_{\gamma}=0.1$ kW/cm$^{2}$ and for absorber thicknesses of 100~nm, 50~nm, and 20~nm, respectively. For the 20~nm cell, the thickness of the doped layers was reduced to 10 nm. The effects of the increasing built-in fields on the characteristics under illumination [Fig.~\ref{fig:td}(b)] are partially masked by the large change in the optical intensity at the different thicknesses, while recombination -- and, hence, dark current -- is strongly enhanced due to transitions between field-induced tail states [Fig.~\ref{fig:td}(a)]. This is demonstrated in Figs.~\ref{fig:absem_td} and \ref{fig:absorpt_td} displaying the absorption and emission spectra and the total absorptance at $V_{\textrm{bias}}=0.8$ V, respectively, for the different absorber thicknesses. There is no substantial variation in the value of the absorption coefficient at the photon energy $E_{\gamma}=1.44$ eV of the monochromatic illumination (indicated by the vertical arrow), however, the corresponding absorptance exhibits the pronounced thickness dependence observed in the short-circuit current in Fig.~\ref{fig:td}(b). This strongly non-monotonous behavior originates in the variation of the optical intensity for the different thickness configurations. The emission spectrum, on the other hand, is red-shifted and broadened with decreasing absorber thickness due to the stronger tailing of the joint density of states with growing built-in field. As a consequence of this field-induced shrinking of the effective band gap, the emission rate obtained from integration of the spectrum is drastically enhanced in the ultra-scaled regime.

\begin{figure}[t]
\begin{center}
\includegraphics[width=0.4\textwidth]{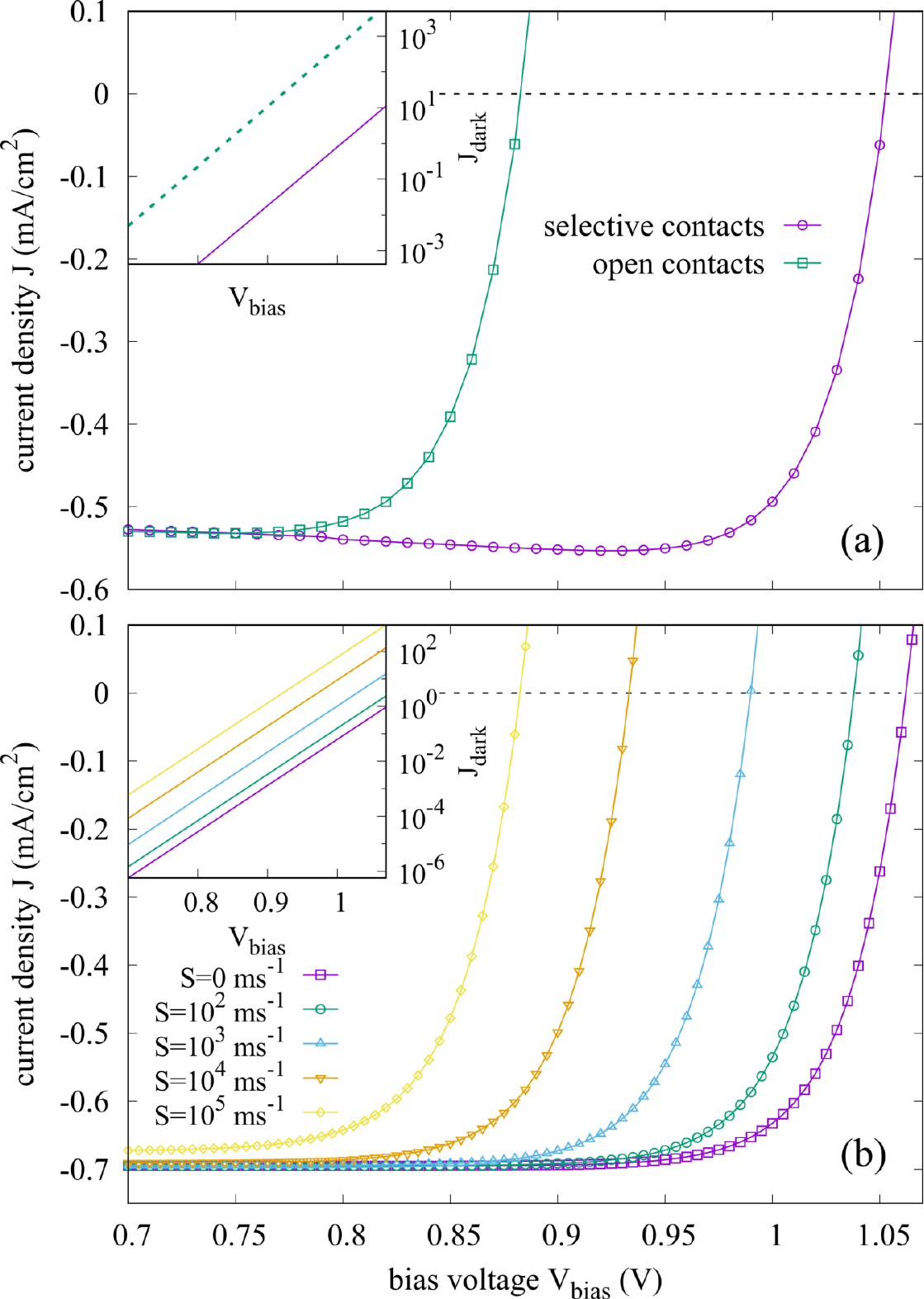}
\caption{(a) JV-characteristics for open and selective contacts for a 100 nm $p$-$i$-$n$ diode under monochromatic illumination with $E_{\gamma}=1.44$ eV at an intensity of 0.1 kW/cm$^2$, as optained using the NEGF formalism. The open contacts lead to severe $V_{\textrm{OC}}$-reduction due to dark leakage currents. The corresponding dark current-voltage characteristics are displayed in the inset (log scale). (b) Semiclassical JV-characteristics under consideration of leakage loss via a corresponding surface recombination term characterized by the surface recombination velocity $S$. The inset displays again the dark current for the different values of $S$.  \label{fig:JV_leakage}}
\end{center}
\end{figure}

\begin{figure}[t]
\begin{center}
\includegraphics[width=0.45\textwidth]{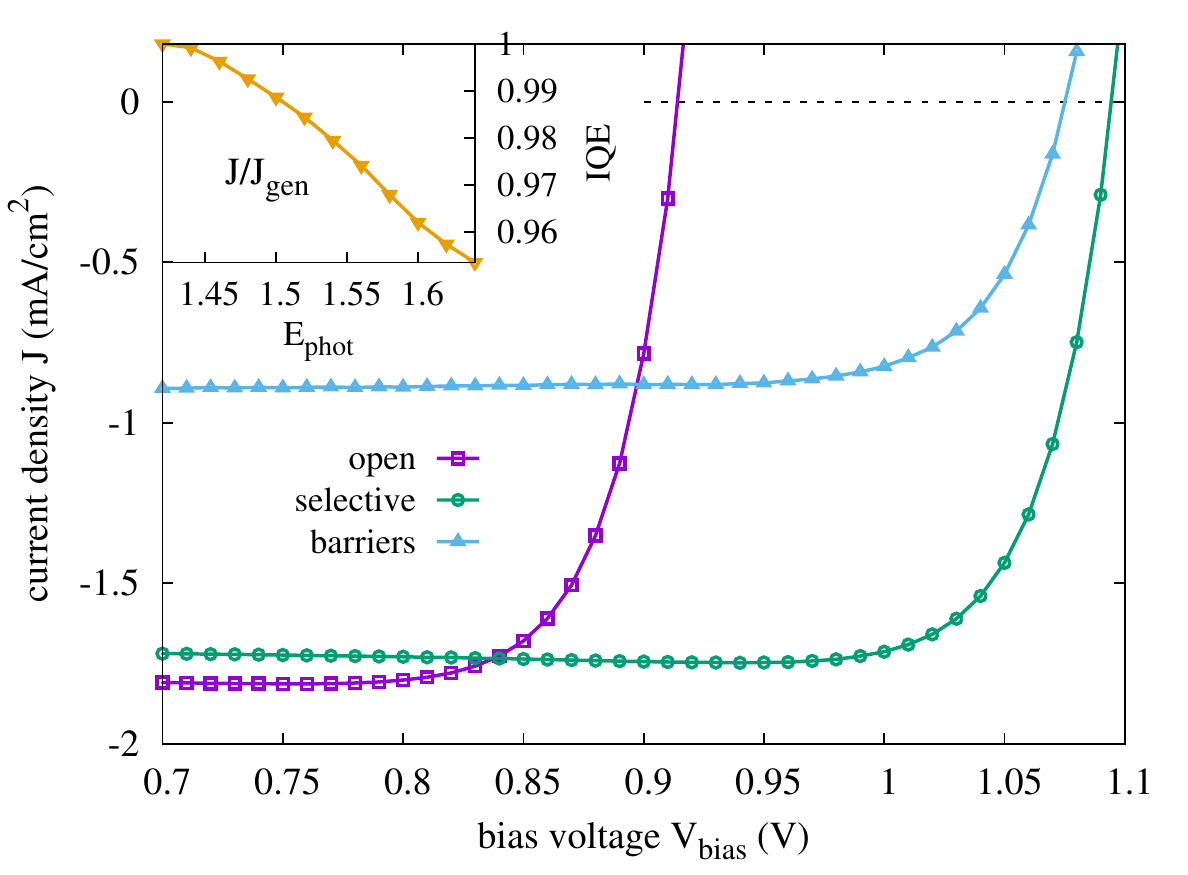}
\caption{NEGF-based JV-characteristics for a 50 nm $p$-$i$-$n$ structure with open contacts ("open"), carrier selective contacts ("selective") and with electron/hole blocking layers ("barriers"). The inset shows the ratio of extracted to generated photocurrent at short circuit conditions, which depends on the photon energy: the farther away from the band edge the carriers are generated, the higher the probability of photocurrent leakage.\label{fig:JV_leak_barr}}
\end{center}
\end{figure} 

\begin{figure}[t]
\begin{center}
\includegraphics[width=0.45\textwidth]{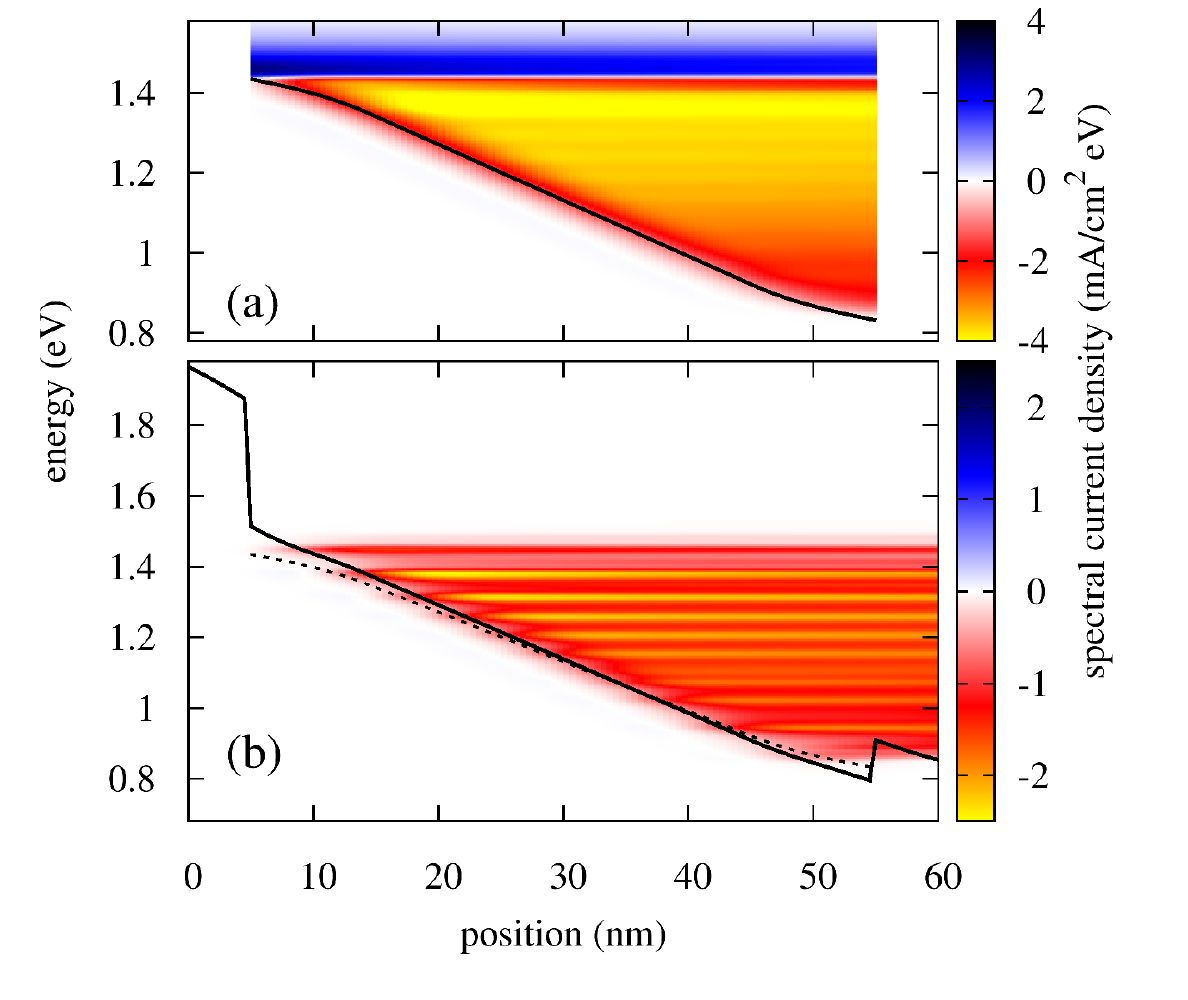}
\caption{Spectral electron current for (a) the 50 nm $p$-$i$-$n$ cell with open contacts at $V_{\textrm{bias}}=0.87$ V, featuring significant leakage of electronically injected carriers, and (b) the architecture with blocking layers, where leakage is prevented, but at the price of sizable reduction and modification of the photocurrent spectrum. The dashed line in (b) corresponds to the band profile from (a) and shows that the built-in field is only weakly modified by the presence of barriers. \label{fig:currspect}}
\end{center}
\end{figure} 

The effect of the contacts is mainly related to their behavior with respect to absorption, reflection and injection of charge carriers. In an ideal solar cell, the contacts are perfectly carrier selective, i.e., they are ideally transmissive for majority carriers, while reflecting all minority carriers, thereby acting as the "semipermeable membrane" that is essential for photovoltaic device operation \cite{wuerfel:book_05}. In the NEGF picture, the contacts are described by corresponding boundary self-energies that relate the absorber states to the extended states of the electrodes \cite{datta:95}.
At vanishing contact barriers ($\Delta E_{\textrm{C,V}}\rightarrow 0$, "open"), there are photovoltaic performance losses due to leakage of both electrically and optically injected carriers. As shown in Fig.~\ref{fig:JV_leakage}(a), the leakage component of the dark current -- i.e., the intraband component -- results in severe reduction of $V_{\textrm{OC}}$. In the semiclassical picture, this behavior can be reproduced by increasing the value of the surface recombination velocity $S$ at minority carrier contacts [Fig.~\ref{fig:JV_leakage}(b)]: for increasing $S$, $V_{\textrm{OC}}$ is gradually reduced, and at very large surface recombination, also $J_{\textrm{SC}}$ is degraded. In the NEGF picture, the reduction of $J_{\textrm{SC}}$ additionally depends on photon energy: the larger the distance to the band edge, the more likely is the extraction of photogenerated carriers to the "wrong" -- i.e., minority carrier -- contact, especially for carriers generated far away from the contact \cite{cavassilas:15}. This behavior is displayed in the inset of Fig.~\ref{fig:JV_leak_barr} showing the ratio $J/J_{\textrm{gen}}$ of  extracted to generated charge current for the 50 nm thick absorber.

\begin{figure}[t] 
\begin{center}
\includegraphics[width=0.5\textwidth]{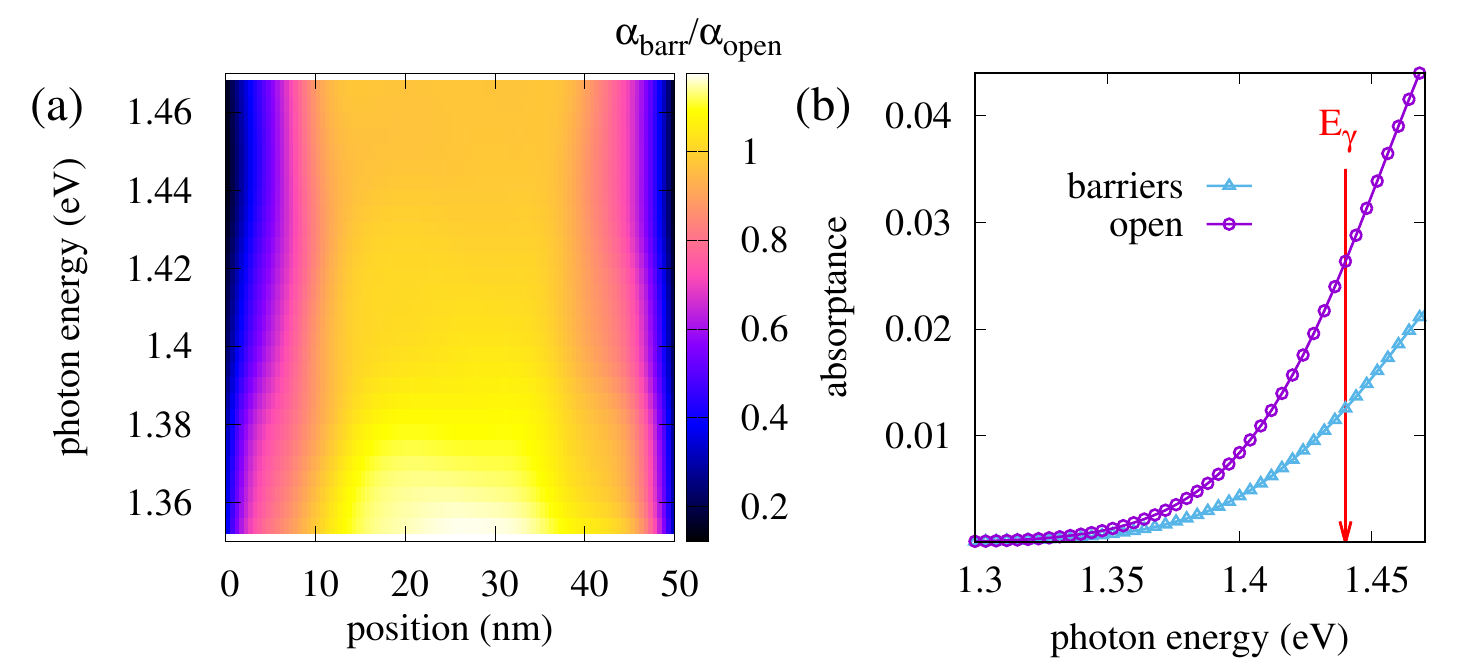}
\caption{(a) Ratio $\alpha_{\textrm{barr}}/\alpha_{\textrm{bulk}}$ of the spatially resolved absorption coefficients for the structure with blocking layers and the diode with open contacts, revealing the strong reduction of absorption in close vicinity of the barriers. (b) Comparison of the overall absorptance of the two structures, which explains the large difference in photocurrent seen in Fig.~\ref{fig:JV_leak_barr}.\label{fig:abs_pin_barr}}
\end{center}
\end{figure}

The leakage losses can be partially mitigated by introducing electron and hole blocking layers (EBL/HBL), as displayed in Fig.~\ref{fig:JV_leak_barr}. In the present case, d$_{\textrm{EBL}}$=5~nm of Al$_{40}$Ga$_{60}$As is chosen for the EBL, and the HBL is composed of d$_{\textrm{HBL}}$=5~nm of In$_{49}$Ga$_{51}$P. While dark leakage current is successfully reduced, there is also a reduction in photogeneration due to the impact of barriers on the electronic states inside the absorber, especially in close proximity to the interfaces, as can be verified in Figs.~\ref{fig:currspect}(a) and (b) displaying the spectral current of electrons in the two architectures. For comparison, the band profile of the open contact cell (a) is also indicated (dashed lines) in subfigure (b) showing the situation for the system with barrier layers. Obviously, the small variation in built-in field cannot explain the observed photocurrent reduction. Indeed, the comparison of the absorption coefficients of the two structures, shown in Fig.~\ref{fig:abs_pin_barr}(a), reveals sizable modifications induced by the presence of confining barriers, most pronounced in the vicinity of the blocking layers, which explains the difference in absorptance displayed in Fig.~\ref{fig:abs_pin_barr}(b).

Thus, while a homogeneous 50~nm GaAs slab behaves essentially bulk-like from an electronic point of view, the modification of contact regions has sizeable impact on the electronic structure and, hence, on the optoelectronic response of the device. As a result, simulation approaches that take into account only optical confinement effects and neglect the electronic modification of the absorber due to built-in fields and contact regions become inappropriate and should be replaced by a more comprehensive and generally valid picture, such as the NEGF framework used here.

In conclusion, the computational investigation of ultra-thin GaAs solar cells based on a rigorous approach that is valid beyond the limits of the semi-classical bulk picture reveals significant discrepancies with respect to the latter due to the non-classical effects that strong built-in fields and contact barriers have on absorption, emission and charge carrier extraction. The NEGF approach presented here will therefore provide a valuable tool for the investigation, design and optimization of solar cell device structures in the ultra-scaled regime, where the exact device geometry and configuration starts to affect not only the optical modes, but also the relevant electronic states of the absorber.

We gratefully acknowledge funding from the European Commission Horizon 2020 project No. 676629 (''EoCoE'') and computing time granted on the supercomputer JURECA at J\"ulich Supercomputing Centre (JSC).

 \balance

\bibliographystyle{aipnum4-1}

\begin{thebibliography}{10}


\bibitem{kayes:11}
B.~Kayes, H.~Nie, R.~Twist, S.~Spruytte, F.~Reinhardt, I.~Kizilyalli, and
  G.~Higashi, ``27.6\% conversion efficiency, a new record for single-junction
  solar cells under 1 sun illumination,'' in \emph{Photovoltaic Specialists
  Conference (PVSC), 2011 37th IEEE}, pp. 000004--000008,  2011.

\bibitem{wang:13_jpv}
X.~Wang, M.~Khan, J.~Gray, M.~Alam, and M.~Lundstrom, ``Design of GaAs solar
  cells operating close to the Shockley-Queisser limit,'' \emph{IEEE J. Photovolt.}, vol.~3, no.~2, pp. 737--744, 2013.

\bibitem{steiner:13}
M.~A. Steiner, J.~F. Geisz, I.~García, D.~J. Friedman, A.~Duda, and S.~R.
  Kurtz, ``Optical enhancement of the open-circuit voltage in high quality GaAs
  solar cells,'' \emph{J. Appl. Phys.}, vol. 113, no.~12, art. no.~123109, 2013.

\bibitem{wang:13}
Z.~Wang, T.~White, and K.~Catchpole, ``Plasmonic near-field enhancement for
  planar ultra-thin photovoltaics,'' \emph{IEEE Photon. J.}, vol.~5,
  no.~5, art. no.~8400608, 2013.

\bibitem{massiot:14}
I.~Massiot, N.~Vandamme, N.~Bardou, C.~Dupuis, A.~Lemaître, J.-F. Guillemoles,
  and S.~Collin, ``Metal nanogrid for broadband multiresonant light-harvesting
  in ultrathin GaAs layers,'' \emph{ACS Photonics}, vol.~1, no.~9, pp.
  878--884, 2014.

\bibitem{yang:14}
W.~Yang, J.~Becker, S.~Liu, Y.-S. Kuo, J.-J. Li, B.~Landini, K.~Campman, and
  Y.-H. Zhang, ``Ultra-thin GaAs single-junction solar cells integrated with a
  reflective back scattering layer,'' \emph{J. Appl. Phys.}, vol.
  115, no.~20, art. no.~203105, 2014. 

\bibitem{vandamme:15}
N.~Vandamme, C.~Hung-Ling, A.~Gaucher, B.~Behaghel, A.~Lemaitre, A.~Cattoni,
  C.~Dupuis, N.~Bardou, J.-F. Guillemoles, and S.~Collin, ``Ultrathin GaAs
  solar cells with a silver back mirror,'' \emph{IEEE J. Photovolt.}, vol.~5, no.~2, pp. 565--570, 2015.


\bibitem{ae:jcel_11}
U.~Aeberhard, ``{Theory and simulation of quantum photovoltaic devices based on
  the non-equilibrium Green’s function formalism},'' \emph{J. Comput.
  Electron.}, vol.~10, pp. 394--413, 2011.

\bibitem{ae:jpe_14}
U.~Aeberhard, ``Simulation of nanostructure-based and ultra-thin film solar
  cell devices beyond the classical picture,'' \emph{J. Photon. Energy},
  vol.~4, no.~1, art. no.~042099, 2014.

\bibitem{ae:oqel_14}
U.~Aeberhard, ``{Photon Green's functions for a consistent theory of absorption
  and emission in nanostructure-based solar cell devices},'' {\em Opt. Quantum.
  Electron.}, vol.~46, pp.~791--796, 2014.  
  
 \bibitem{ae:jpv_16}
U.~Aeberhard, ``Simulation of Ultrathin Solar Cells Beyond the Limits of the Semiclassical Bulk Picture,'' \emph{IEEE J. Photovolt.}, vol.~6, no.~3, pp. 654--660, 2016. 

\bibitem{roosbroeck:54}
W.~Van~Roosbroeck and W.~Shockley, ``Photon-radiative recombination of
  electrons and holes in germanium,'' \emph{Phys. Rev.}, vol.~94, art. no.~1558, 1954.

\bibitem{ae:prb_08}
U.~Aeberhard and R.~H. Morf, ``Microscopic nonequilibrium theory of quantum
  well solar cells,'' {\em Phys. Rev. B}, vol.~77, art. no.~125343, 2008.

  
\bibitem{lake:97}
R.~Lake, G.~Klimeck, R.~Bowen, and D.~Jovanovic, ``Single and multiband
  modelling of quantum electron transport through layered semiconductor
  devices,'' \emph{J. Appl. Phys.}, vol.~81, art. no.~7845, 1997.
  
\bibitem{ae:prb89_14}
U.~Aeberhard, ``Quantum-kinetic theory of steady-state photocurrent generation in
  thin films: Coherent versus incoherent coupling,'' \emph{Phys. Rev. B},
  vol.~89, art. no.~115303, 2014.

\bibitem{wuerfel:book_05}
P.~W\"urfel, {\em Physics of Solar Cells}.
\newblock Wiley-VCH, 2005.

\bibitem{datta:95}
S.~Datta, {\em Electronic Transport in Mesoscopic Systems}.
\newblock Cambridge University Press, 1995.

\bibitem{cavassilas:15}
N.~Cavassilas, C.~Gelly, F.~Michelini, and M.~Bescond, ``Reflective barrier
  optimization in ultrathin single-junction GaAs solar cell,''
  \emph{IEEE J. Photovolt.}, vol.~5, no.~6, pp. 1621--1625, 2015.




\end{thebibliography}

\end{document}